\documentclass[twocolumn,showpacs,preprintnumbers,amsmath,amssymb,superscriptaddress]{revtex4}

\usepackage{graphicx}
\usepackage{dcolumn}
\usepackage{bm}

\usepackage{slashbox}

\makeatletter
\def\simleq{\mathrel{\mathpalette\gl@align<}}
\def\simgeq{\mathrel{\mathpalette\gl@align>}}
\def\gl@align#1#2{\lower.6ex\vbox{\baselineskip\z@skip\lineskip\z@
     \ialign{$\m@th#1\hfill##\hfil$\crcr#2\crcr\sim\crcr}}}
\makeatother

\newcommand{\gf}{\gamma_5}

\newcommand{\nn}{\nonumber\\}

\newcommand{\fslash}[1]{\ooalign{\hfil/\hfil\crcr$#1$}}
\newcommand{\bra}{\langle}
\newcommand{\ket}{\rangle}
\newcommand{\braket}[1]{\bra #1 \ket}

\begin{document}

\preprint{UK/09-01}

\title{
Nucleon strangeness form factors from $N_f=2+1$
clover fermion lattice QCD
}

\author{Takumi Doi} 
 \email{doi@pa.uky.edu}
 \altaffiliation[\\Present Address:]{
Graduate School of Pure and Applied Science,
University of Tsukuba,
Tennodai 1-1-1,
Tsukuba, Ibaraki 305-8571, Japan.}
\author{Mridupawan Deka}%
\author{Shao-Jing Dong}%
\author{Terrence Draper}%
\author{Keh-Fei Liu}%
\author{Devdatta Mankame}%
\affiliation{%
Department of Physics and Astronomy,
University of Kentucky, Lexington KY 40506, USA
}%

\author{Nilmani Mathur}
\affiliation{%
Department of Theoretical Physics,
Tata Institute of Fundamental Research,
Mumbai 40005, India
}%

\author{Thomas Streuer}
\affiliation{%
Institute for Theoretical Physics,
University of Regensburg, 93040 Regensburg, Germany
}%

\collaboration{$\chi$QCD Collaboration}
\noaffiliation


\begin{abstract}

We present the $N_f=2+1$ clover fermion lattice QCD calculation
of the nucleon strangeness form factors.
We evaluate disconnected insertions using the Z(4) stochastic method,
along with unbiased subtractions from the hopping parameter expansion.
We find that increasing the number of
nucleon sources for each configuration improves the signal significantly.
We obtain $G_M^s(0) = -0.017(25)(07)$, where the first error is statistical,
and the second is the uncertainties in $Q^2$ and chiral extrapolations.
This is consistent with experimental values,
and has an order of magnitude smaller error.

\end{abstract}

\maketitle

\section{Introduction}
\label{sec:intro}

The structure of the nucleon plays an essential role
in understanding the dynamics of QCD,
and 
experiments have been providing 
many intriguing and unexpected results 
for decades.
%
%
%
%
%
%
In particular, the strangeness content of the nucleon 
attracts a great deal of interest lately.
As the lightest non-valence quark structure,
it is an ideal probe
for
the virtual sea quarks in the nucleon.
Extensive
experimental/theoretical studies indicate that
the strangeness content varies significantly depending 
on the quantum number carried by the $s\bar{s}$ pair:
the 
scalar density 
is about $10$--$20$\% 
of that of up, down quarks,
the quark spin is $-10$ to $0$\%
of the nucleon,
and 
the momentum fraction is only 
a few percent of the nucleon.
%
%
%
%
Recently, 
intensive experiments
have been carried out
for the electromagnetic form factors
by
SAMPLE~\cite{sample.04}, 
A4~\cite{a4}, 
HAPPEX~\cite{happex}, and
G0~\cite{g0.05}, 
through parity-violating electron scattering (PVES).
The global analyses~\cite{young06,j.liu07,pate08} 
have produced, e.g.,
$G_E^s(Q^2) = -0.008(16)$ and
$G_M^s(Q^2) = 0.29(21)$ at $Q^2 = 0.1 {\rm GeV}^2$~\cite{j.liu07},
but substantial errors still exist so that the 
results are consistent with zero.
%
%
%
%
Making tighter constraints on these form factors from the theoretical side 
is one of the challenges in QCD calculation.
Moreover, such 
constraints, together with experimental inputs,
can lead to more precise determinations of various interesting quantities,
such as the axial form factor $G_A^s$~\cite{pate08},
and the electroweak radiative corrections including 
the nucleon anapole moment, $\tilde{G}_A$~\cite{PAVI02:musolf,young06}.

\begin{figure}[bt]
\begin{center}
%
\includegraphics[width=0.2\textwidth]{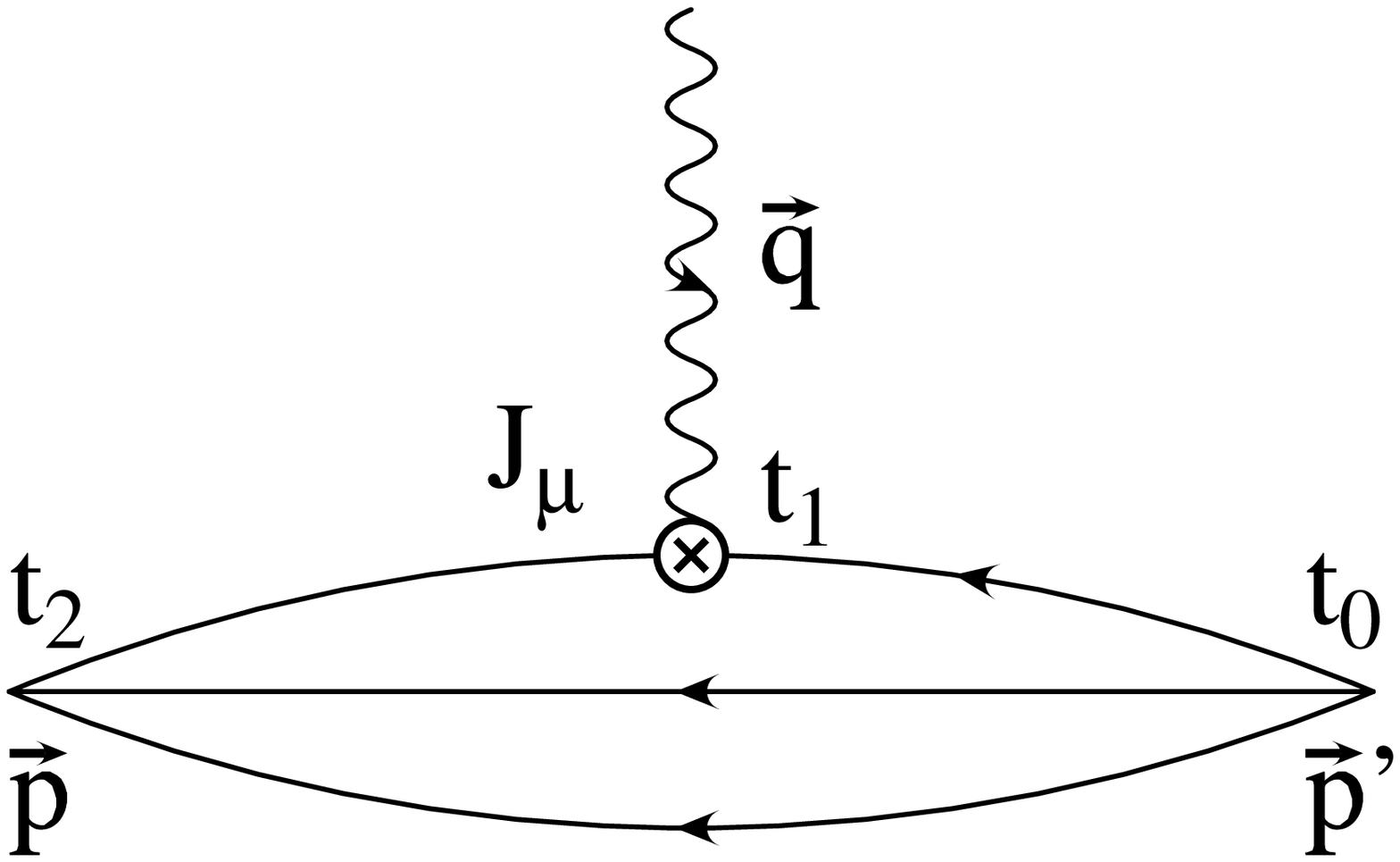}
\hspace*{1em}
\includegraphics[width=0.2\textwidth]{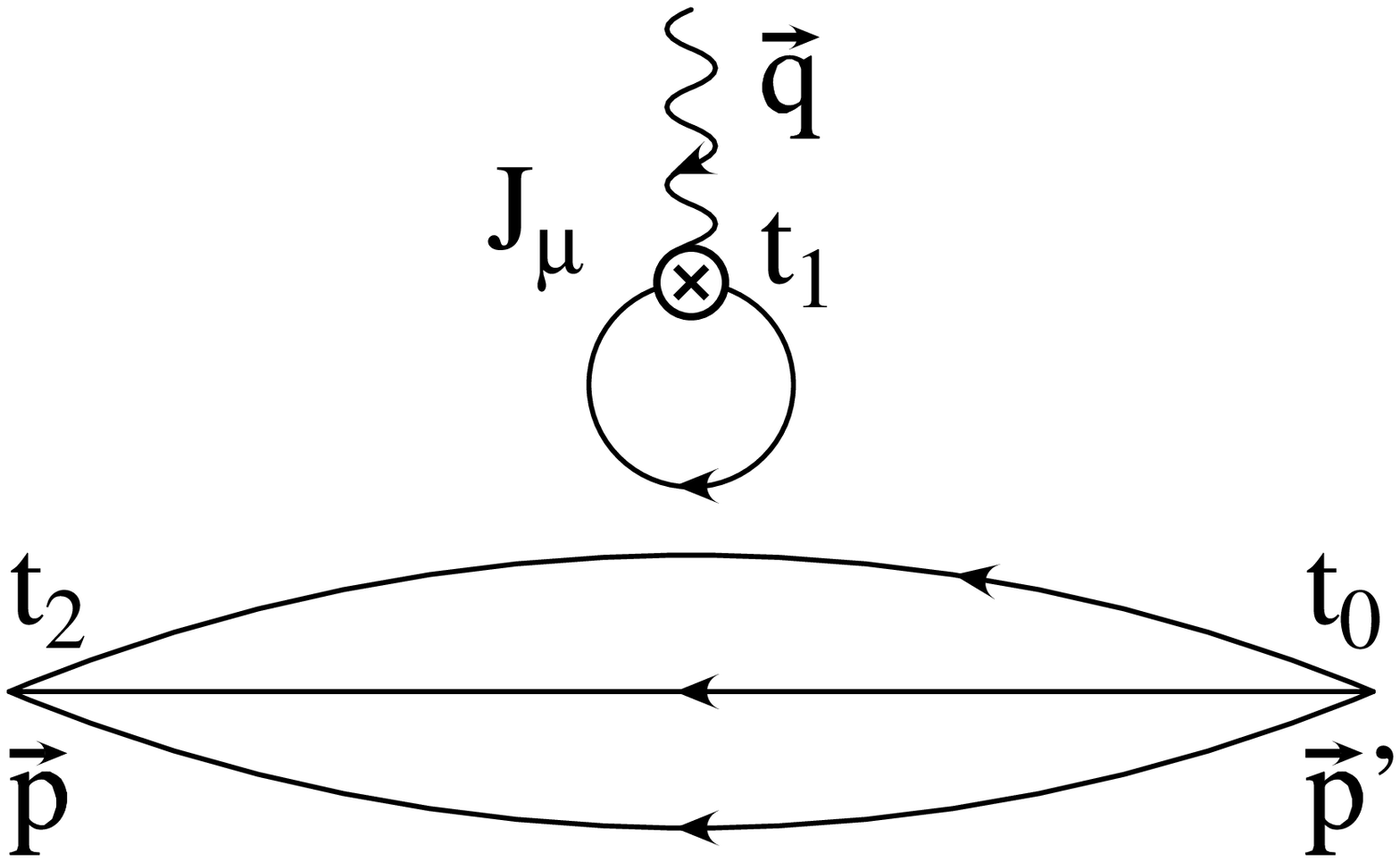}
\end{center}
\vspace*{-4mm}
\caption{
\label{fig:3pt_diag}
Two different 
representations 
for the 3pt 
function.
}
\end{figure}

Unfortunately,
the theoretical status of strangeness form factors 
remains quite uncertain. 
%
%
For instance,
the values for the magnetic moment 
$G_M^s(0)$ 
from different analyses vary widely:
$-0.31(9)$ 
in the dispersion relation (DR) with pole ansatz~\cite{jaffe89},
$-(0.15-0.51)$
in DR with scattering kaon clouds~\cite{hammer99},
$0.035$ 
in the quark model~\cite{geiger97},
%
%
%
and
$0.08-0.13$
in the chiral quark-soliton model~\cite{silva06}.
%
%
%
The analyses for the electric form factor are
similarly ambiguous.
%
%
%
Under these circumstances, the most desirable study
is the first-principle calculation in QCD, such as the lattice simulation.
However, the calculation requires
the evaluation of the 
disconnected insertion (DI) (Fig.~\ref{fig:3pt_diag} (right)),
which is 
much more difficult 
compared to the connected insertion (CI) (Fig.~\ref{fig:3pt_diag} (left)) calculation,
because the straightforward calculation of DI requires all-to-all propagators,
and is prohibitively expensive.
%
Consequently, there are only few DI calculations, where
the all-to-all propagators are stochastically estimated.
The first calculation was done in the quenched approximation with Wilson fermion~\cite{smm:ky_quenched}
and gave $G_M^s(0) = -0.36(20)$, and $-0.28(10)$ from the updated calculation~\cite{smm:ky_quenched2}.
Another quenched DI calculation with Wilson fermion~\cite{smm:randy}
obtained $G_M^s(0.1 {\rm GeV}^2)= 0.05(6)$.
There are also several indirect estimates
using quenched~\cite{smm:lein,wang08} or unquenched~\cite{huey07}
lattice data for the CI part,
and
the experimental magnetic moments (or electric charge radii) for octet baryons
as inputs under the assumption of isospin symmetry.
These estimates obtained
$G_M^s(0) = -0.046(19)$~\cite{smm:lein} and
$G_M^s(0) = -0.066(26)$~\cite{huey07}.

In this paper,
we provide the
first full QCD lattice simulation of the direct DI calculation
with high statistics.
The paper is organized as follows.
In Sec.~\ref{sec:formalism}, the formulation and parameter setup 
of the lattice calculation
are presented. Technical details to improve the DI evaluation 
are also given.
In Sec.~\ref{sec:results}, we present the lattice QCD data,
and carefully examine the possible systematic uncertainties.
Sec.~\ref{sec:summary} is devoted to the summary of our results.
%
%

\section{Formalism}
\label{sec:formalism}

We employ 
$N_f=2+1$ dynamical 
configurations
with nonperturbatively ${\cal O}(a)$ improved clover fermion 
and 
RG-improved gauge action
generated by CP-PACS/JLQCD Collaborations~\cite{conf:tsukuba2+1}.
We use 
$\beta=1.83$ and
$c_{sw}=1.7610$ 
configurations 
with the lattice size of $L^3 \times T = 16^3\times 32$.
The lattice spacing was determined
using 
$K$-input or
$\phi$-input~\cite{conf:tsukuba2+1}.
We use the averaged scale of
$a^{-1} = 1.625 {\rm GeV}$ hereafter,
as the uncertainty of the scale is negligible
compared to the statistical error in our results.
For the hopping parameters of $u$,$d$ quarks ($\kappa_{ud}$) and
$s$ quark ($\kappa_s$), we use
$\kappa_{ud} =$ $0.13825$, $0.13800$, and $0.13760$,
which correspond to $m_\pi =$ $0.60$, $0.70$, and $0.84$ ${\rm GeV}$,
respectively, and $\kappa_s = 0.13760$ is fixed.
We perform the calculation only at the dynamical quark mass points,
with the periodic boundary condition 
in all 
directions.


\begin{table}
\caption{\label{tab:mass}
The setup 
parameters as well as 
hadron masses. 
$N_{noise}$ is the number of 
noises in the stochastic estimate
and $N_{src}$ is the number of 
nucleon sources
for each configuration.
}
\begin{ruledtabular}
\begin{tabular}{lcccccc}
$\kappa_{ud}$ & $N_{conf}$ & $N_{noise}$ & $N_{src}$ & $m_\pi a$ & $m_K a$ & $m_N a$ \\
\hline
0.13760 & 800 & 600 & 64 & 0.5141(5) & 0.5141(5) & 1.0859(12) \\
0.13800 & 810 & 600 & 82 & 0.4302(6) & 0.4540(5) & 0.9623(16) \\
0.13825 & 810 & 800 & 82 & 0.3717(7) & 0.4141(6) & 0.8844(20) \\
\end{tabular}
\end{ruledtabular}
\end{table}

We calculate the three point function (3pt) $\Pi^{\rm 3pt}_{J_\mu}$
and two point function (2pt) $\Pi^{\rm 2pt}$ defined as
%
%
\begin{eqnarray}
\lefteqn{
\Pi^{\rm 3pt}_{J_\mu}(\vec{p},t_2;\ \vec{q},t_1;\ \vec{p'}=\vec{p}-\vec{q},t_0)
= 
\sum_{\vec{x_2},\vec{x_1}}
e^{-i\vec{p}\cdot(\vec{x}_2-\vec{x}_0)}
\times
}
\nn
&&
\hspace*{-6mm}
e^{+i\vec{q}\cdot(\vec{x}_1-\vec{x}_0)}
\braket{0|{\rm T}\left[
\chi_N(\vec{x}_2,t_2) {J_\mu}(\vec{x}_1,t_1) \bar{\chi}_N(\vec{x}_0,t_0)
\right] |0} 
\label{eq:3pt} \\[4mm]
\lefteqn{
\Pi^{\rm 2pt}(\vec{p},t;\ t_0) 
}
\nn
&=& 
\sum_{\vec{x}} e^{-i\vec{p}\cdot(\vec{x}-\vec{x}_0)}
\braket{0|{\rm T}\left[
\chi_N(\vec{x},t) \bar{\chi}_N(\vec{x}_0,t_0)
\right] |0} ,
\label{eq:2pt}
%
%
\end{eqnarray}
where 
%
%
$\chi_N = \epsilon_{abc} (u_a^T C\gf d_b) u_c$ 
is the nucleon interpolation field 
and the insertion 
$J_\mu$ is given by the point-split conserved current
%
$
J_\mu (x+ \mu/2)
= 
(1/2) \times
$
{\small
$
\left[
\bar{s}(x+\mu) (1+\gamma_\mu) U_\mu^\dag(x) s(x)
- 
\bar{s}(x) (1-\gamma_\mu) U_\mu(x) s(x+\mu)
\right] .
\label{eq:vec}
$
}
%

Electromagnetic form factors can be obtained 
using $\vec{p}=\vec{0}$, $\vec{p'}= -\vec{q}$ kinematics 
for the forward propagation ($t_2 \gg t_1 \gg t_0$)~\cite{smm:ky_quenched}.
In this work, we consider the backward propagation ($t_2 \ll t_1 \ll t_0$) as well,
in order to increase 
statistics. The formulas for 
Sachs electric (magnetic) form factors
$G_E^s$ ($G_M^s$) 
are summarized as
%
%
\begin{eqnarray}
%
%
%
%
R_\mu^{\pm} (\Gamma_{\rm pol}^\pm) &\equiv&
\frac{
{\rm Tr}\left[
\Gamma_{\rm pol}^\pm\cdot
\Pi^{\rm 3pt}_{J_{\mu}}(\vec{0},{t}_2;\ \pm\vec{q},{t}_1;\ -\vec{q},t_0)
\right]
}
{
{\rm Tr} \left[
\Gamma_e^\pm\cdot \Pi^{\rm 2pt}_{}(\pm\vec{q},t_1;\ t_0)
\right]
}
%
\nn
&&
%
\qquad
\times
\frac{
{\rm Tr} \left[
\Gamma_e^\pm\cdot \Pi^{\rm 2pt}_{}(\vec{0},t_1;\ t_0)
\right]
}
{
{\rm Tr} \left[
\Gamma_e^\pm\cdot \Pi^{\rm 2pt}_{}(\vec{0},t_2;\ t_0)
\right]
} ,
\label{eq:ratio,kin1,fb}
%
%
%
\end{eqnarray}
%
%
\begin{eqnarray}
R_{\mu=4}^{\pm}(\Gamma_{\rm pol}^\pm=\Gamma_e^\pm) &=& \pm G_E^s(Q^2), 
\label{eq:ele1,kin1,fb} \\ 
R_{\mu=i}^{\pm}(\Gamma_{\rm pol}^\pm=\Gamma_k^\pm) &=& 
\frac{\mp \epsilon_{ijk} q_j}{E^q_N+m_N}
G_M^s(Q^2) ,
\label{eq:mag,kin1,fb}
%
%
\end{eqnarray}
where 
$\{i,j,k\} \neq 4$,
$
\Gamma_e^\pm \equiv (1\pm\gamma_4)/2
$ ,
$
\Gamma_k^\pm \equiv (\pm i)/2 \times (1\pm\gamma_4) \gf \gamma_k
$ 
and
%
$E^q_N \equiv \sqrt{m_N^2 + \vec{q}^{\,2}}$.
%
%
%
The upper (lower) sign corresponds to 
the forward (backward) propagation.

Furthermore, we consider 
another kinematics of
$\vec{p}=\vec{q}$, $\vec{p'}=\vec{0}$,
%
%
%
where the analogs of 
Eqs.~(\ref{eq:ratio,kin1,fb}),
(\ref{eq:ele1,kin1,fb}), 
(\ref{eq:mag,kin1,fb}) hold.
We find that the results from 
the latter kinematics
have similar size of statistical errors as those from
the former,
and the average of them yields better results.
Hereafter, we present results from total average of 
two kinematics and forward/backward propagations,
unless otherwise noted.
%

%
\begin{figure}[bt]
\begin{center}
\includegraphics[width=0.3\textwidth,angle=270]{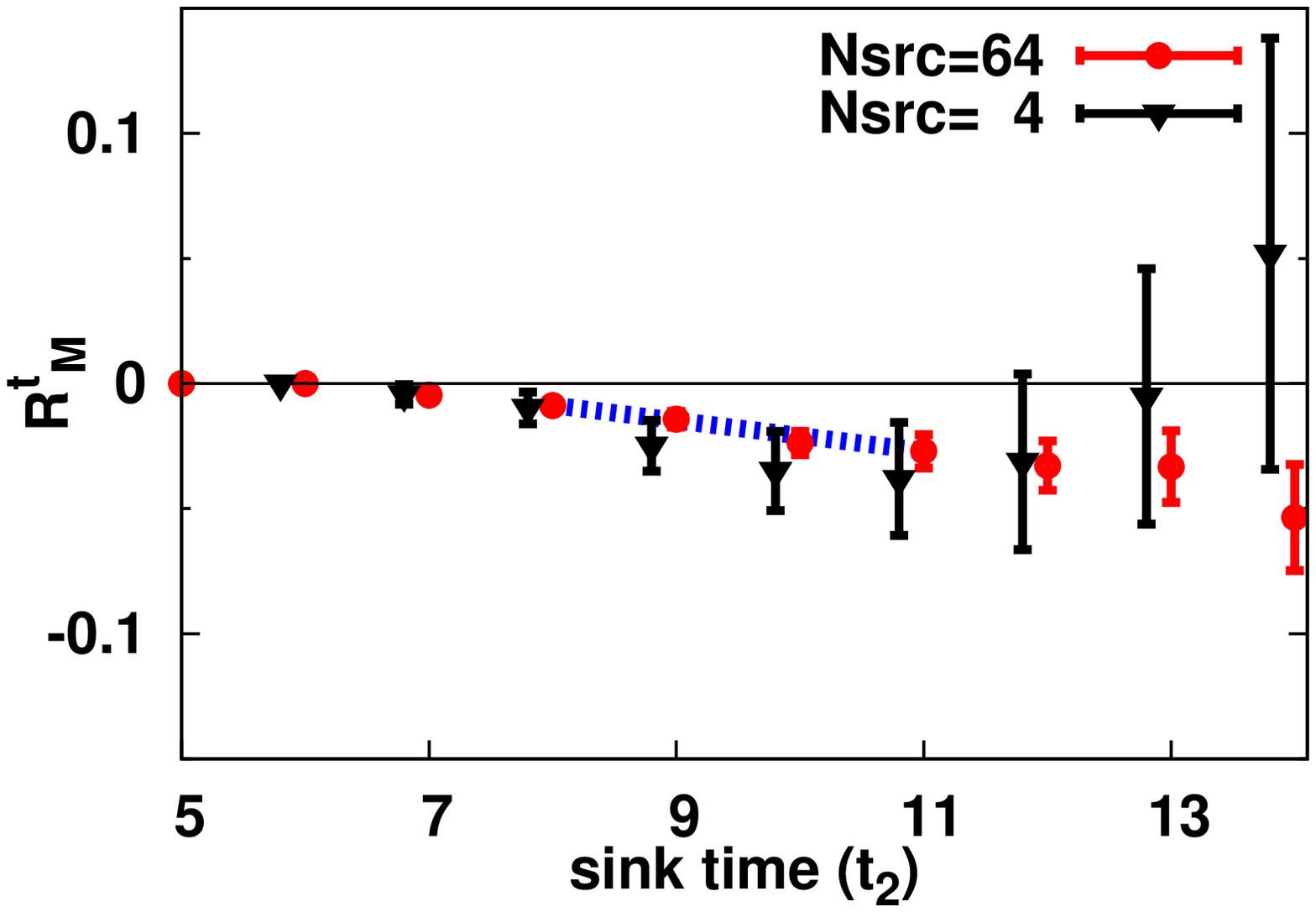}
\includegraphics[width=0.3\textwidth,angle=270]{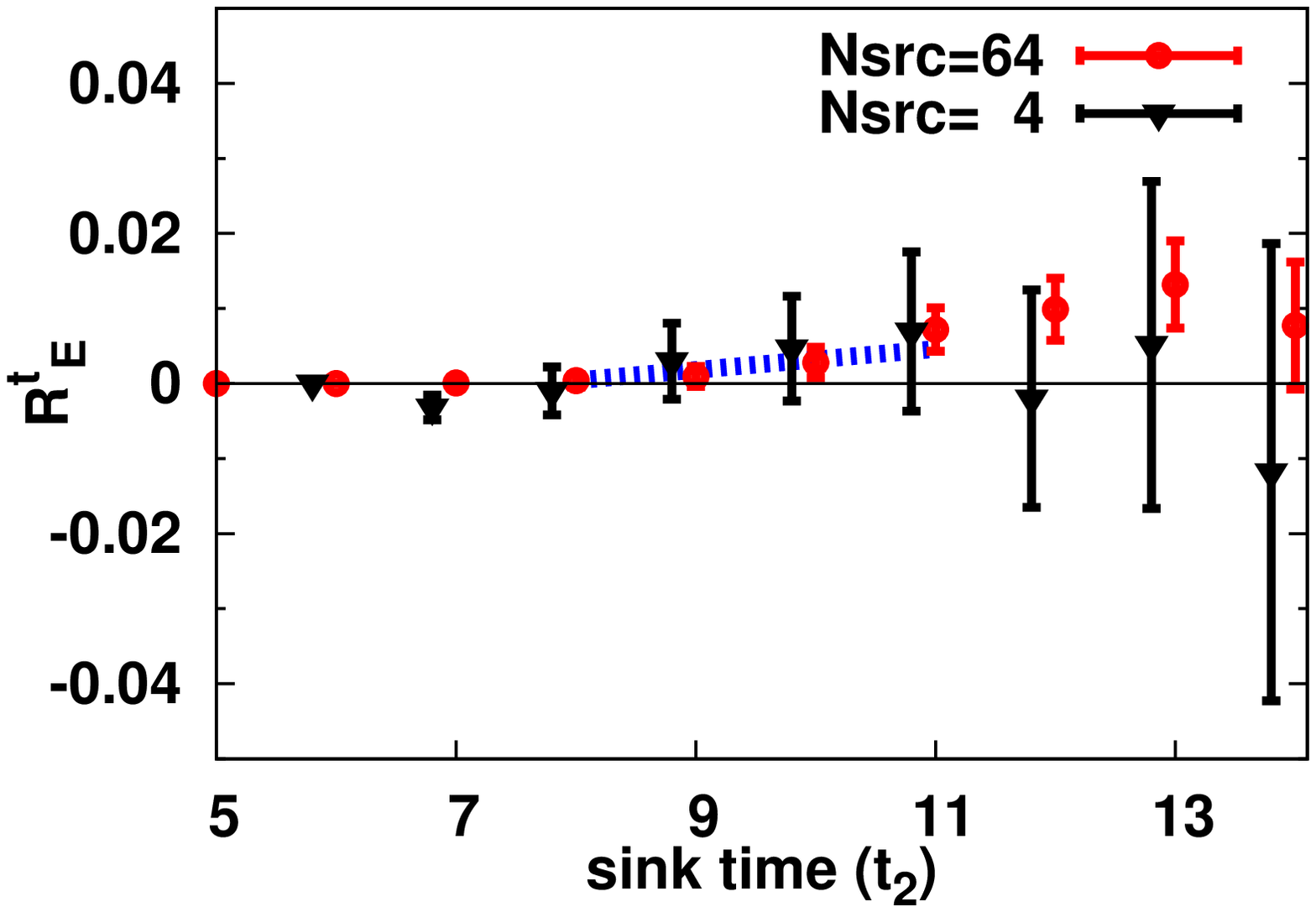}
\end{center}
\vspace*{-8mm}
\caption{
 \label{fig:tsum}
$R_M^t$ (upper) and $R_E^t$ (lower)
with 
$\kappa_{ud}=0.13760$, 
$\vec{q}^{\,2}=2\cdot (2\pi/La)^2$,
$N_{src}=64$ (circles) and $N_{src}=4$ 
(triangles, with offset for visibility),
plotted against the nucleon sink time $t_2$.
The dashed line is the linear fit
where the slope corresponds to the form factor.
}
\end{figure}
\begin{table*}[tbh]
\caption{\label{tab:s_em}
Lattice results for the strangeness electromagnetic form factors
with the momentum-squared $\vec{q}^{\,2} = n\cdot ( 2\pi/ La)^2$.
}
\begin{ruledtabular}
\begin{tabular}{cccccccccc}
                & \multicolumn{4}{c}{$G_M^s(Q^2) (\times 10^{-2})$} & \multicolumn{5}{c}{$G_E^s(Q^2) (\times 10^{-2})$} \\
$\kappa_{ud} \backslash n$ & 1 & 2 & 3 & 4                   & 0 & 1 & 2 & 3 & 4 \\ \hline
0.13760 & -0.96(29) & -0.61(17) & -0.57(20) & -0.20(25)      &  0.21(21)  &  0.01(10) & 0.14(08) & 0.22(11) & 0.10(15) \\
0.13800 & -0.76(37) & -0.76(24) & -0.57(32) &  0.04(41)      &  0.15(28)  &  0.16(14) & 0.14(12) & 0.15(19) & 0.46(29) \\
0.13825 & -1.09(41) & -1.02(27) & -0.67(33) & -0.25(47)      & -0.04(33)  & -0.16(15) & 0.36(13) & 0.27(20) & 0.71(31)
\end{tabular}
\end{ruledtabular}
\end{table*}
\begin{table}
\caption{\label{tab:Q2fit}
The parameters fitted against $Q^2$ behavior. 
}
\begin{ruledtabular}
\begin{tabular}{ccccccc}
              & \multicolumn{3}{c}{$G_M^s(Q^2)$}              & \multicolumn{2}{c}{$G_E^s(Q^2)$} \\
$\kappa_{ud}$ & $G_M^s(0)$         & $\Lambda a$ & $\chi^2/{\rm dof}$ & $g_E^s$            & $\chi^2/{\rm dof}$ \\
              & $(\times 10^{-2})$ &             &                    & $(\times 10^{-2})$ &                    \\
\hline
0.13760 & -1.7(12) & 0.66(29) & 0.34(83)      & 1.2(5) & 0.56(89) \\
0.13800 & -1.4(09) & 0.77(40) & 0.77(126)     & 1.4(6) & 0.34(70) \\
0.13825 & -1.9(11) & 0.80(40) & 0.38(87)      & 1.9(7) & 2.68(189)
\end{tabular}
\end{ruledtabular}
\end{table}

The calculations of 3pt 
functions
for the strangeness current
need
the evaluation of DI.
We use the stochastic method~\cite{DI:noise},
with Z(4) noises in color, spin and space-time indices.
We generate independent noises for different configurations,
in order to avoid possible auto-correlation.
%
%
To reduce fluctuations, 
we use the charge conjugation and $\gamma_5$-hermiticity (CH), and parity symmetry.
For instance, we find that the information for the $G_M^s$ is coded
in the product of ${\rm Re}(\Pi^{\rm 2pt}) \times {\rm Re}({\rm loop})$,
and filtering out the imaginary parts reduces the noises~\cite{angular:ky_quenched,x:deka}.
%
%
We also perform unbiased subtractions~\cite{DI:hpe} to 
reduce
the off-diagonal contaminations to the variance.
For subtraction operators,
we employ those obtained through
hopping parameter expansion (HPE) for the propagator $M^{-1}$~\cite{DI:hpe},
%
$
\frac{1}{2\kappa} M^{-1} = \frac{1}{1+C} + \frac{1}{1+C} (\kappa D) \frac{1}{1+C} 
+ \cdots
%
$
%
where $D$ denotes the Wilson-Dirac operator and $C$ the clover term.
We subtract up to order $(\kappa D)^4$ term,
and observe that 
the statistical error is 
reduced
by a factor of 2.

In the stochastic method, 
it is quite expensive
to achieve a good signal to noise ratio (S/N) just by increasing $N_{noise}$
because S/N improves with $\sqrt{N_{noise}}$.
In view of this, we use many nucleon point sources $N_{src}$ 
in the evaluation of the 2pt part for each configuration~\cite{x:deka}.
Since the calculations of the loop part and 2pt part
are independent of each other, this is expected to be an efficient way. 
In particular, for the $N_{noise} \gg N_{src}$ case,
we observe that S/N improves almost ideally, by a factor of $\sqrt{N_{src}}$.
We take $N_{src}=64$ for $\kappa_{ud} = 0.13760$
and $N_{src}=82$ for $\kappa_{ud} = 0.13800, 0.13825$, 
where locations of sources are taken 
so that they are 
separated
in 4D-volume as much as possible.
The 
calculation parameters 
as well as basic hadron masses are tabulated in 
Tab.~\ref{tab:mass}.

There are several ways~\cite{smm:ky_quenched,smm:randy} to extract the matrix elements from 
Eqs.~(\ref{eq:ele1,kin1,fb}) and (\ref{eq:mag,kin1,fb})
with various $t_1, t_2$ results.
Among them, it is advocated~\cite{smm:ky_quenched,smm:ky_quenched2,angular:ky_quenched,x:deka} to 
take the 
summation over the insertion time $t_1$,
symbolically 
given as
%
$
R^{t}_{E,M} \equiv
\frac{1}{K_{E,M}^\pm}
\sum_{t_1=t_0+t_{s}}^{t_2-t_{s}}
R^{\pm}_\mu
=
{\rm const.} + t_2 \times G_{E,M}^s ,
\label{eq:tsum}
$
%
where $K_{E,M}^\pm$ are trivial kinematic factors 
appearing in
Eqs.~(\ref{eq:ele1,kin1,fb}) and (\ref{eq:mag,kin1,fb}),
and $t_s$ is chosen so that the error is minimal.
We thus obtain $G_{E,M}^s$ as 
the linear slope of $R^{t}_{E,M}$ against $t_2$.
In order to achieve ground state saturation,
we use the data only for $t_2 \geq 7$~\cite{x:deka}.


\section{Results and Discussions}
\label{sec:results}

We calculate 
for the 
five smallest momentum-squared points,
%
$
\vec{q}^{\,2} = n\cdot ( 2\pi/ La)^2\ 
$
$(n=0$--$4)$, 
%
%
which correspond to 
$|\vec{q}| = 0, 0.64, 0.90, 1.1, 1.3$ GeV.
%
%
%
Typical figures for 
$R^{t}_{M}$, $R^{t}_{E}$ 
are shown in Fig.~\ref{fig:tsum}.
One can 
observe 
the significant S/N improvement by increasing $N_{src}$.
%
The numerical results are given
in Tab.~\ref{tab:s_em}.
We note 
$G_E^s(0)$ 
are consistent with zero, 
which serves as a test of the calculations.
Of particular interest is that,
for all $\kappa_{ud}$ simulations,
$G_M^s(Q^2)$ is found 
to be negative with 
2-3 
$\sigma$ signals for low $Q^2$ regions.
%
%
%
%

In order to determine the magnetic moment, 
the $Q^2$ dependence of $G_M^s(Q^2)$ is studied.
We employ the dipole form in the $Q^2$ fit,
$G_M^s(Q^2) = G_M^s(0) / (1+Q^2/\Lambda^2)^2$,
where reasonable agreement with lattice data is observed.
%
%
%
For the electric form factor,
we employ 
$G_E^s(Q^2) = g_E^s \cdot Q^2 / (1+Q^2/\Lambda^2)^2$,
considering that $G_E^s(0)=0$ from the vector current 
conservation.
In the practical fit of $G_E^s(Q^2)$, however, 
reliable extraction of the pole mass $\Lambda$ is impossible
because $G_E^s(Q^2)$ data are almost zero within error.
Therefore, we assume that $G_E^s(Q^2)$ 
has the same pole mass as $G_M^s(Q^2)$, 
and perform a one-parameter fit for $g_E^s$.
The obtained parameters are given in Tab.~\ref{tab:Q2fit}.
We also test the simultaneous fit of 
$G_M^s(Q^2)$ and $G_E^s(Q^2)$ with 
three parameters of $G_M^s(0), \Lambda, g_E^s$, 
and confirm that the results are consistent with the values
in Tab.~\ref{tab:Q2fit}.

Finally, we perform the chiral extrapolation for the fitted parameters.
Since our quark masses are relatively heavy, 
we consider only the leading dependence on $m_K$,
which is obtained by heavy baryon chiral perturbation theory (HB$\chi$PT). 
For the magnetic moment $G_M^s(0)$, we fit linearly in terms of $m_K$~\cite{chPT:hemmert,smm:ky_quenched}.
For the pole mass $\Lambda$, we take that the magnetic mean-square radius
$\braket{r^2_s}_M \equiv -6 \frac{d G_M^s}{d Q^2}|_{Q^2=0} = 12 G_M^s(0) / \Lambda^2$
behaves as $1/m_K$~\cite{chPT:hemmert}.
For $g_E^s$, we use the electric radius
$\braket{r^2_s}_E \equiv -6 \frac{d G_E^s}{d Q^2}|_{Q^2=0} = -6 g_E^s$
which has an $\ln (m_K/\mu)$ behavior~\cite{chPT:hemmert}, and we take the scale $\mu=1$ GeV.
The chiral extrapolated results are
$G_M^s(0) = -0.017(25)$,
$\Lambda a = 0.58(16)$,
$\braket{r_s^2}_M = -7.4(71) \times 10^{-3} {\rm fm}^2$
and
$g_E^s = 0.027(16)$
(or $\braket{r_s^2}_E = -2.4(15) \times 10^{-3} {\rm fm}^2$).


%
%

%
%
%
\begin{figure}[tb]
\begin{center}
%
\includegraphics[width=0.28\textwidth,angle=270]{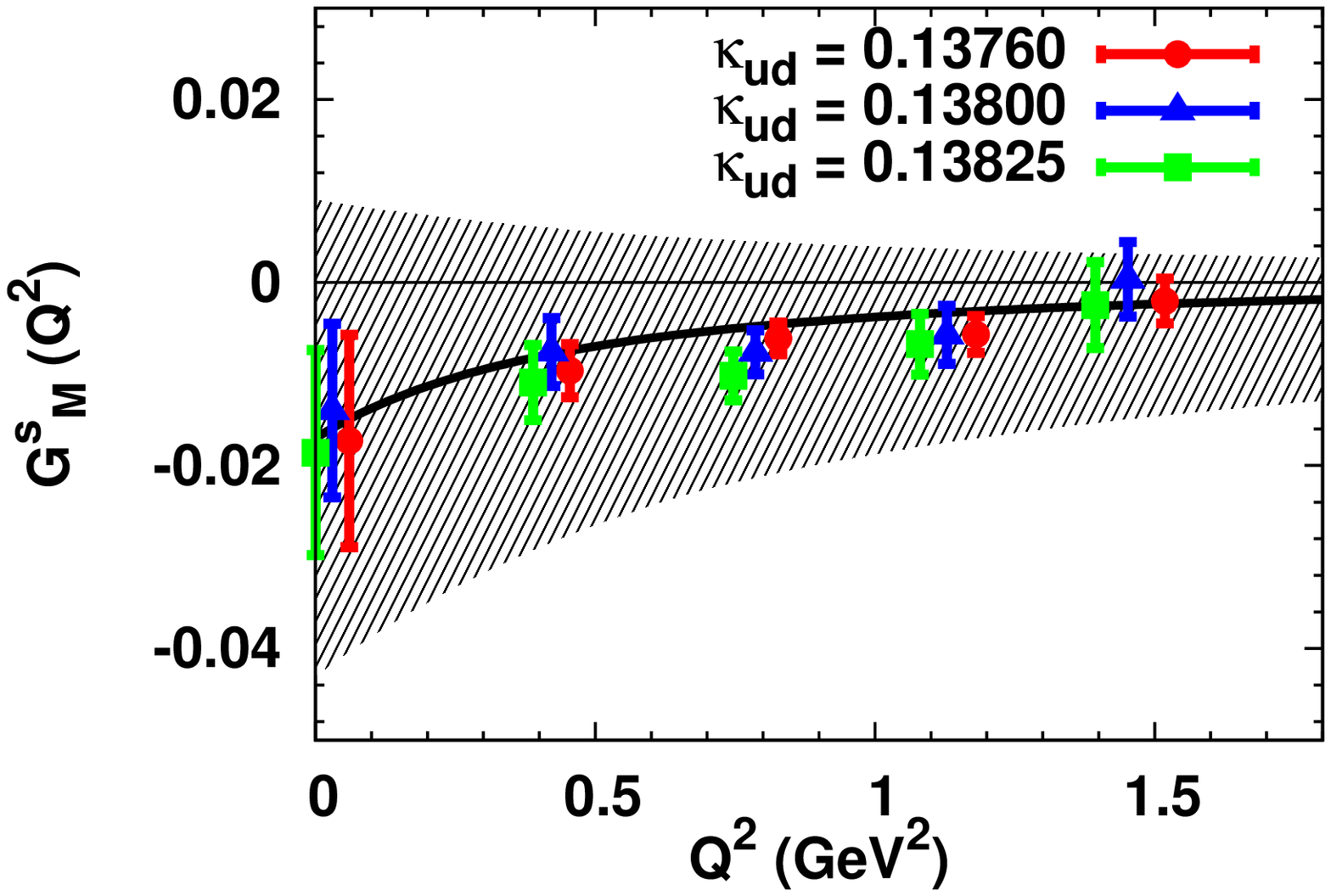} \\[-3mm]
\includegraphics[width=0.28\textwidth,angle=270]{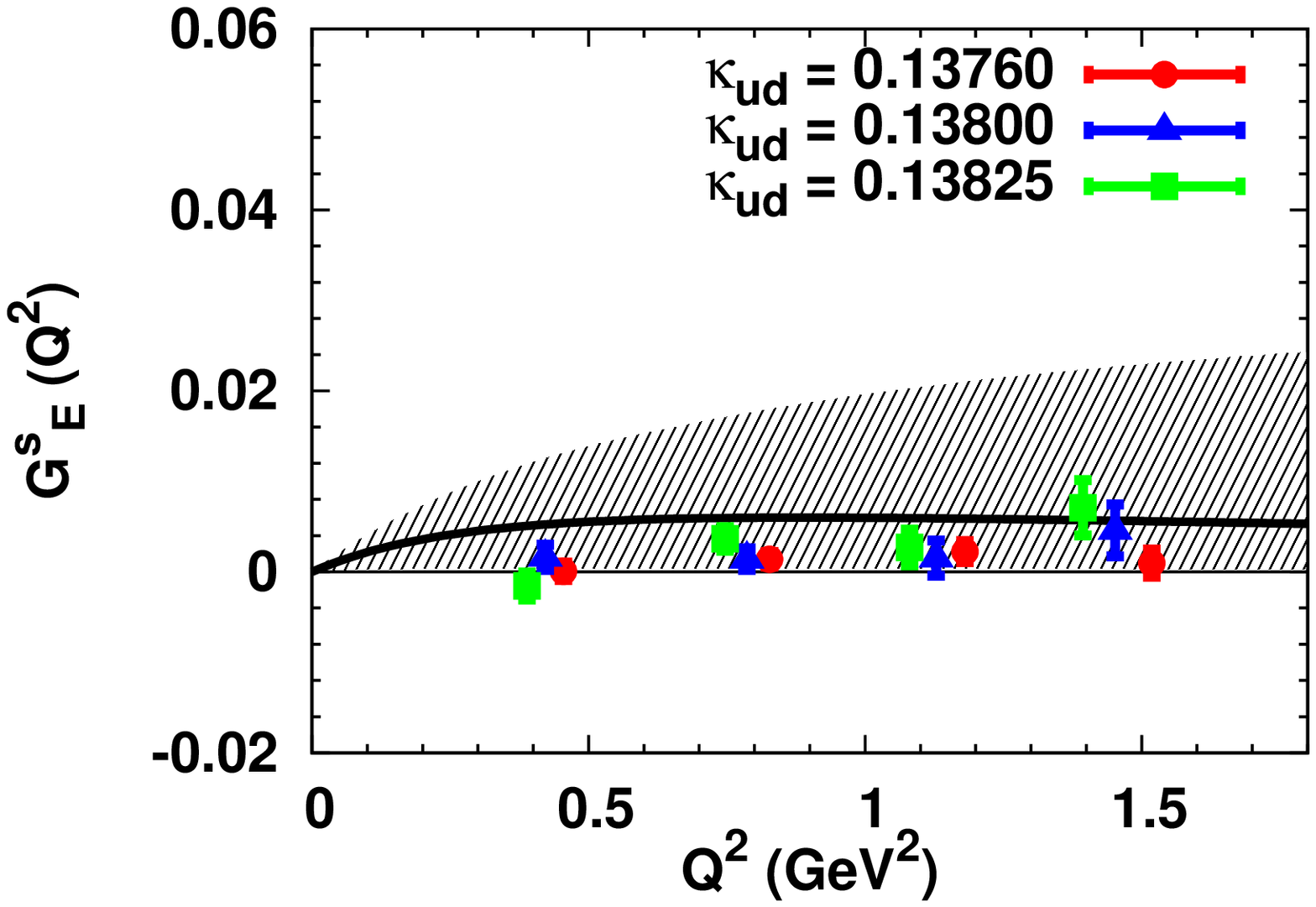}
%
\end{center}
\caption{
 \label{fig:res}
The chiral extrapolated results for $G_M^s(Q^2)$ (upper) 
and $G_E^s(Q^2)$ (lower) plotted with solid lines.
Shaded regions 
represent
the error-band 
with statistical and systematic error
added in quadrature.
Shown
together are the lattice data 
(and $Q^2$-extrapolated $G_M^s(0)$)
for
$\kappa_{ud}=$ 
0.13760 (circles),
0.13800 (triangles),
0.13825 (squares)
with offset for visibility.
}
\end{figure}

Before quoting the final results,
we consider the systematic 
uncertainties yet to be addressed.
First, we analyze the ambiguity of $Q^2$ dependence in form factors,
which is rather unknown for the strangeness.
We also study the monopole form, $G_M^s(Q^2) = G_M^s(0) / (1+Q^2/\tilde{\Lambda}^2)$,
$G_E^s(Q^2) = g_E^s\cdot Q^2 / (1+Q^2/\tilde{\Lambda}^2)$,
and find that the precision of the lattice data cannot 
disentangle the difference of the dipole/monopole behaviors.
The results from the monopole fit are
$G_M^s(0) = -0.024(39)$,
$\tilde{\Lambda} a = 0.34(17)$,
$\braket{r_s^2}_M = -12(18) \times 10^{-3} {\rm fm}^2$
and
$g_E^s = 0.035(17)$
(or $\braket{r_s^2}_E = -3.1(15) \times 10^{-3} {\rm fm}^2$),
which are consistent with those from the dipole fit
and have relatively larger statistical errors.

Second, we study the uncertainties in chiral extrapolation
by testing two alternative extrapolations.
In the first one,
we take into account 
the nucleon mass dependence on the quark mass,
using the lattice nucleon mass. 
From the physical viewpoint,
this corresponds to measuring the magnetic moment 
not in units of 
lattice 
magneton but physical
magneton~\cite{qcdsf:formfactor}.
We obtain
$G_M^s(0) = -0.017(24)$, 
$\Lambda a = 0.77(21)$,
$\braket{r_s^2}_M = -4.6(43) \times 10^{-3} {\rm fm}^2$
from the dipole fit, and
$G_M^s(0) = -0.023(36)$, 
$\tilde{\Lambda} a = 0.45(22)$,
$\braket{r_s^2}_M = -7(11) \times 10^{-3} {\rm fm}^2$
from the monopole fit,
%
while $G_E^s$ cannot be fitted
because 
fit dof are not sufficient.
Note that results are consistent with previous analyses.
In the second alternative,
we use the linear fit 
in terms of 
$m_K^2$, 
observing the results have weak quark mass dependence.
We obtain
$G_M^s(0) = -0.017(21)$, 
$\braket{r_s^2}_M = -3(17) \times 10^{-3} {\rm fm}^2$,
$g_E^s = 0.023(12)$
(or $\braket{r_s^2}_E = -2.0(10) \times 10^{-3} {\rm fm}^2$)
from the dipole fit, and
$G_M^s(0) = -0.023(33)$, 
$\braket{r_s^2}_M = -8(43) \times 10^{-3} {\rm fm}^2$,
$g_E^s = 0.029(12)$
(or $\braket{r_s^2}_E = -2.5(11) \times 10^{-3} {\rm fm}^2$)
from the monopole fit,
while pole masses are found to suffer from too large statistical errors
to extract useful information.
We find again the results are consistent with previous analyses.
While a further clarification with physically light quark mass simulation
and a check on convergence of HB$\chi$PT~\cite{chPT:hammer}
is desirable, 
we use the dependence of results on different extrapolations
as systematic uncertainties in the chiral extrapolation.

Third, we examine the contamination from excited states.
Because our spectroscopy study 
indicates that the mass of Roper resonance is
massive compared to the $S_{11}$ state
on the current lattice~\cite{ky:meng},
the dominant contaminations are (transition) form factors 
associated with $S_{11}$.
%
%
On this point, we find that 
such contaminations can be eliminated theoretically,
by making the following substitutions: 
%
$
\Gamma_e^\pm \rightarrow
\tilde{\Gamma}_e^\pm \equiv
\left( 1 \pm {m_{N^*}}/{E_{N^*}^q}\cdot \gamma_4 \right)/2
$
%
in Eq.~(\ref{eq:ratio,kin1,fb}),
%
%
%
%
$
\{ \Gamma_e^\pm,\ \Gamma_k^\pm \} \rightarrow
\{
\tilde{\Gamma}_\pm^{p'_-} \Gamma_e^\pm,\ \tilde{\Gamma}_\pm^{p'_-} \Gamma_k^\pm \}
$
%
in Eqs.~(\ref{eq:ele1,kin1,fb}) and (\ref{eq:mag,kin1,fb}).
%
%
%
%
%
%
%
%
Here, 
%
%
%
$m_{N^*}$ denotes the $S_{11}$ mass, 
$E_{N^*}^q \equiv \sqrt{m_{N^*}^2 + \vec{q}^{\,2}}$,
%
$
\tilde{\Gamma}_\pm^{p'_-}
\equiv ({m_{N^*} \mp i\fslash{p}'_-})/({2m_{N^*}})
$
%
with
$p'_- \equiv (E_{N^*}^{p'}, \vec{p'})$.
Note also that the (r.h.s.) of 
Eqs.~(\ref{eq:ele1,kin1,fb}) and (\ref{eq:mag,kin1,fb})
have modifications in the kinematical factor,
which we do not show explicitly.
%
%
It is found that 
the results from these equations are basically the same as before,
so we conclude that the contamination regarding the $S_{11}$ state is 
negligible.


As remaining sources of systematic error,
one might worry that 
the finite volume artifact could be substantial considering that 
the physical spacial size of the lattice is about $(2 {\rm fm})^3$.
However,
we recall that 
Sachs radii are found to be quite small, 
$ |\braket{r_s^2}_{E,M}| \ll 0.1 {\rm fm}^2$,
which 
indicates a small finite volume artifact.
For the discretization error,
we first examine the error associated with the finite momentum $\vec{q}$,
using the dispersion relation of the nucleon.
We find that the nucleon energy at each $\vec{q}$ 
on the lattice
is consistent with 
the dispersion relation, and conclude that finite $(qa)$ discretization
error is negligible.
As another discretization error,
we note that $m_N$ ($m_K$) is found to have 6 (8) \% error for the current 
configurations~\cite{conf:tsukuba2+1,conf:tsukuba2+1:pre}. 
Considering the dependence of $G_{E,M}^s$ on these masses,
we estimate that the discretization errors in our results 
amount to $\simleq 10$\%, and are much smaller than the statistical errors.
Of course, more quantitative investigations on 
these issues 
are necessary 
with larger and finer lattices,
and such work is in progress.

Here, we present our final results.
For the magnetic moment, 
$G_M^s(0) = -0.017(25)(07)$, 
where the first error is statistical and
the second is systematic 
from uncertainties of the $Q^2$ extrapolation and chiral extrapolation.
We also obtain $\Lambda a = 0.58(16)(19)$ for dipole mass
or $\tilde{\Lambda} a = 0.34(17)(11)$ for monopole mass,
and $g_E^s = 0.027(16)(08)$.
These lead to 
$G_M^s(Q^2) = -0.015(23)$,
$G_E^s(Q^2) =  0.0022(19)$ at $Q^2=0.1 {\rm GeV}^2$,
where error is obtained by quadrature from 
statistical and systematic errors.
%
Note that 
these are consistent with the world averaged data~\cite{young06,j.liu07,pate08},
with an order of magnitude smaller error.
In Fig.~\ref{fig:res}, 
we plot 
$G_M^s(Q^2)$, $G_E^s(Q^2)$,
where the shaded regions 
correspond to the square-summed error.
The lattice data for each $\kappa_{ud}$
are also plotted.

\section{Summary}
\label{sec:summary}

We have studied
the strangeness electromagnetic form factors of the nucleon
using $N_f=2+1$ clover fermion configurations.
In the evaluation of the disconnected insertion (DI),
the Z(4) stochastic method along with unbiased subtractions
from the hopping parameter expansion has been used.
We have developed several techniques to achieve a good S/N
in the DI calculation,
and found that 
increasing the number of nucleon sources for each configuration
is particularly efficient.
For all quark mass simulations,
$G_M^s(Q^2)$ has been found 
to be negative with 
2-3 
$\sigma$ signals for low $Q^2$ regions.
Upon $Q^2$ and chiral extrapolations,
we have obtained
$G_M^s(Q^2=0) = -0.017(25)(07)$,
where the first error is statistical,
and the second reflects the uncertainties in
$Q^2$
and chiral extrapolations.
We have also obtained 
$G_M^s(Q^2) = -0.015(23)$,
$G_E^s(Q^2) =  0.0022(19)$ at $Q^2=0.1 {\rm GeV}^2$,
where error is obtained by quadrature from 
statistical and systematic errors.
These results are consistent with experimental values,
and our errors are an order of magnitude smaller.

\begin{acknowledgments}
We thank the CP-PACS/JLQCD Collaborations
for their configurations.
This work was supported  in part by 
U.S. DOE grant DE-FG05-84ER40154.
Research of N.M. is supported by Ramanujan Fellowship.
The calculation was performed 
at Jefferson Lab, Fermilab
and 
the Univ. of Kentucky,
partly using the Chroma Library~\cite{chroma}.
\end{acknowledgments}

\vspace*{-2mm}


\end{document}